\documentclass[journal]{new-aiaa}

\usepackage{amsmath}
\usepackage{bm}
\usepackage{graphicx}
\usepackage{booktabs}
\usepackage{multirow}
\usepackage{xcolor}

\title{Rapid Sequence Cost Evaluation for Multi-Spacecraft Rendezvous: A Permutation-Invariant Set Transformer}

\author{An-yi Huang\footnote{Associate Professor, Orbit Control Team; hay04@foxmail.com.}}
\affil{State Key Laboratory of Astronautic Dynamics, Xi'an, 710043, China}

\begin{document}
\maketitle

\begin{abstract}
Multi-spacecraft multi-rendezvous mission design requires repeated evaluations of the total cost of visiting a target set within a given time window, and the efficiency of this evaluation is the primary barrier to global search. Existing methods either optimize the order and epochs of each candidate sequence, which is expensive, or estimate the cost leg by leg, in which case the error accumulates with the sequence length. This paper proposes a deep model that regresses the total cost of an unordered target set directly: the pairwise transfer cost surface of each target pair is compressed into a latent feature, and a permutation-invariant transformer encoder regresses the sequence cost from these features and the window parameters, without ordering information. Trained on about 0.85 million labeled sequences, the model attains a mean relative error near 5\% and a sub-millisecond evaluation time, about five orders of magnitude faster than precise optimization. Embedded into an evolutionary global search and applied to the GTOC9 problem, it yields the best known solution at every fleet size considered, from seven to ten spacecraft: directly, it produces near-optimal solutions cheaply, and as the initial guess of a differential-evolution refinement it improves the best published solutions further. The evaluator is not specific to this problem or to debris removal: it requires only a pairwise transfer cost function and a sequence-level cost metric monotone in the leg costs, and therefore applies equally to the servicing of satellites in low Earth orbit and to multi-target visits in deep space.
\end{abstract}

\section{Introduction}

SPACE debris has become a major threat to the safety of satellites, and active debris removal missions, in which one or more orbit transfer vehicles rendezvous with and de-orbit multiple debris objects, have been extensively studied in recent years \cite{izzo2015evolving,jpl2018gtoc9,elfman2018trento}. A representative formulation of such missions is the multi-spacecraft multi-rendezvous trajectory optimization problem, where a fleet of spacecraft must visit a large set of targets distributed over neighboring orbits within a limited mission duration, such as the problem of the 9th Global Trajectory Optimization Competition (GTOC9), in which 123 debris objects in low Earth orbits are to be removed by multiple orbit transfer vehicles \cite{jpl2018gtoc9}.

The essential difficulty of this problem lies in the coupling between the discrete and continuous dimensions. For each spacecraft, the visiting order of the assigned targets and the rendezvous epochs are strongly coupled: the transfer cost between two targets varies significantly with the departure epoch, and a favorable order under one time assignment may become unacceptable under another. This dual requirement multiplies the solution space by orders of magnitude, transforming a purely combinatorial problem into one with embedded continuous-time dimensions.

Existing approaches to this problem can be categorized into two main types: (1) heuristic global search methods, in which sequences are constructed and improved by beam search, evolutionary algorithms, or tree search, and the cost of each candidate sequence is evaluated by trajectory optimization \cite{simoes2017beam,tsirogiannis2012graph,zhang2024global,keplerian2026tsp}; and (2) machine-learning-assisted methods, in which neural networks are trained to approximate the cost of a single transfer between two targets and the sequence is then optimized by an external search with the learned pairwise costs \cite{hennes2016fast,zhu2019fast,li2020deep,guo2023dnn,zhang2026pretrained}. More recently, pointer networks and reinforcement learning have been applied to directly generate the visiting order of targets \cite{tomanek2025pointer,rivera2025neural,xu2023active}. However, for the first type, the per-sequence optimization dominates the computational burden and the search scale is strictly limited; for the second type, the pairwise estimation error accumulates along the sequence as the number of legs increases, and the learned sequencing policies require predicting the order, whose learning is known to be expensive and unreliable because the order is strongly coupled with the time assignment.

Note that in the global search stage, the visiting order and the rendezvous epochs are only intermediate products, while the actual quantity consumed by the search is merely the total cost of each candidate sequence. To our knowledge, few studies have attempted to directly evaluate the total cost of a target set without determining the rendezvous order, although this order-free evaluation is precisely what a global search requires.

In the authors' previous studies, an intermediate-orbit-based strategy was proposed to approximately evaluate the cost of a multi-rendezvous sequence without the order information, by rewriting the transfer cost model such that the rendezvous epochs decouple from the visiting order \cite{huang2024orbit,huang2025cost}; and the multi-spacecraft problem was reformulated as a bin-packing problem such that the global search assigns targets to spacecraft without fixing the intra-sequence order \cite{huang2026binpacking}. These analytical approximation models, however, are constrained by their fixed functional forms, and their mean relative errors remain at the level of 10\%. Therefore, a data-driven approach is adopted in this study, in which the sequence cost is regressed directly by a deep network, so that the approximation accuracy is no longer bounded by a preassumed functional form.

The major contributions of this study are threefold. First, an autoencoder-based compression model is established for the pairwise transfer cost profile, which maps the cost--time curve of each target pair into a low-dimensional latent feature. Second, a permutation-invariant deep network is proposed to regress the total cost of an unordered target set directly from the aggregated pairwise features, which obviates the necessity for intra-sequence order determination. Third, the proposed model is embedded into the evolutionary global search framework of \cite{huang2026binpacking} as the sequence evaluator and is applied to the problem of the GTOC9. The verified direct outputs of the search reach 800.4, 713.5, 677.3, and 702.8 MEUR for seven to ten spacecraft, respectively, the ten-spacecraft result being better than the GTOC9 winning score of 731 MEUR \cite{jpl2018gtoc9}, showing that high-quality solutions are now obtained with a high probability rather than as rare outliers; a subsequent differential-evolution refinement stage improves the seven-, eight-, and nine-spacecraft results to 737.8, 676.1, and 660.1 MEUR, and improves the best published nine-spacecraft solution to 629.8 MEUR when the search is warm-started from it; all reported solutions are verified by precise trajectory optimization. The evaluator itself is not specific to this problem: it requires only a pairwise transfer cost function and a sequence-level cost metric that increases with the leg costs, so that the same model can be applied to other multi-target missions, such as the servicing of satellites in low Earth orbit or multi-target visits in deep space.

The remainder of this study is organized as follows: Section~\ref{sec:problem} formulates the sequence cost evaluation problem that dominates the computational cost of multi-spacecraft mission design, together with the mission-level cost model and the scope of the proposed approach. Section~\ref{sec:network} presents the two-layer cost evaluation network and the training dataset generation. Section~\ref{sec:results} reports the training results of both network layers and applies the model to the global search of the GTOC9 problem. Finally, Section~\ref{sec:conclusions} draws the conclusions.

\section{Problem Description}
\label{sec:problem}

The multi-spacecraft multi-rendezvous problem considered in this paper has been formulated in \cite{huang2022global,huang2026binpacking}, and its full statement is not repeated here. Instead, this section isolates the subproblem that dominates the computational cost of such missions and that is the subject of this paper, and it fixes the notation used throughout.

\subsection{Sequence Cost Evaluation Without Order Information}
\label{sec:seqcost}

Consider a set $\mathcal{S} = \{s_1, s_2, \dots, s_N\}$ of $N$ target orbits and a spacecraft that must rendezvous with all of them in succession, starting no earlier than $t_s$ and completing no later than $t_e$. Let $\Delta v(i, j, t, t')$ denote the minimum velocity increment of the transfer that departs from target $i$ at epoch $t$ and arrives at target $j$ at epoch $t'$, evaluated with the transfer model of the mission. The optimal cost of the sequence formed by this target set is
\begin{equation}
\label{eq:task}
\Delta v^*(\mathcal{S}, t_s, t_e) = \min_{\pi,\, \bm{t}} \; \sum_{j=1}^{N-1} \Delta v\!\left(s_{\pi_j},\, s_{\pi_{j+1}},\, t_j,\, t_{j+1}\right),
\end{equation}
where $\pi$ denotes a permutation of $\{1,2,\dots,N\}$, and $\bm{t} = \{t_j\}$ denotes the rendezvous epochs, subject to $t_1 \ge t_s$, $t_N \le t_e$, and the stay-time constraint $t_{j+1} - t_j \ge \tau_{\rm stay}$ imposed by the mission at every target.

Equation~(\ref{eq:task}) is the innermost function of the global optimization algorithms used for this class of problems: candidate target sets are generated and modified iteratively, and each candidate has to be evaluated before it can be compared or selected \cite{huang2025cost,huang2026binpacking,zhang2024global}. A search campaign therefore calls it on the order of $10^7$ times, whereas a single evaluation by numerical optimization of Eq.~(\ref{eq:task}) costs minutes of CPU time even for one sequence; this evaluation bottleneck, rather than the search strategy itself, is what limits the scale of the missions that can be treated. The same quantity is also the one requested in fast mission analysis, where the cost of a candidate target set must be returned interactively without solving a trajectory optimization problem. The task of this study is to construct a model
\begin{equation}
\label{eq:model}
\widehat{\Delta v} = F(\mathcal{S}, t_s, t_e) \approx \Delta v^*(\mathcal{S}, t_s, t_e)
\end{equation}
that replaces Eq.~(\ref{eq:task}) in both usages. Two requirements follow from the way the model is consumed. First, the evaluation must be fast, about one millisecond per sequence, so that the $10^7$ calls of a search campaign become affordable. Second, the model must accept the target set as an unordered collection: the visiting order $\pi$ and the epochs $\bm{t}$ are intermediate products that the outer search never uses, and predicting them is neither necessary nor reliable, since the order is strongly coupled with the time assignment.

\subsection{Mission-Level Cost Model}
\label{sec:missioncost}

The sequence evaluator returns a velocity increment, whereas the objective of a multi-spacecraft mission is a cost. The mapping between the two is recalled here because it is used in Sec.~\ref{sec:scale} to interpret the accuracy of the estimator. If $N_{\rm sc}$ spacecraft divide $N_{\rm tot}$ targets among themselves, the $k$-th spacecraft is assigned an ordered target sequence $\{s_j^k\}$ ($j = 1,2,\dots,N_k$) with rendezvous epochs $\{t_j^k\}$, where $N_k$ is the number of targets of that spacecraft, and its velocity increment is the sum of the transfer increments along the sequence,
\begin{equation}
\label{eq:dvk}
\Delta v_k = \sum_{j=1}^{N_k-1} \Delta v\!\left(s_j^k,\, s_{j+1}^k,\, t_j^k,\, t_{j+1}^k\right).
\end{equation}
The required wet mass follows the rocket equation backward along the sequence,
\begin{equation}
\label{eq:massrec}
\begin{aligned}
m_{N_k} &= m_{\rm dry} + m_{\rm pl}, \\
m_{j} &= m_{j+1}\exp\!\left(\frac{\Delta v_j}{I_{\rm sp} g_0}\right) + m_{\rm pl}, \quad j = N_k-1,\dots,1,
\end{aligned}
\end{equation}
where $\Delta v_j$ is the increment of the $j$-th leg defined in Eq.~(\ref{eq:dvk}), $I_{\rm sp}$ is the specific impulse, $g_0$ the standard gravity, $m_{\rm dry}$ the dry mass left at the end of the mission, and $m_{\rm pl}$ the payload released at each target. The cost of the $k$-th spacecraft is a monotonically increasing function of its total velocity increment,
\begin{equation}
\label{eq:costk}
J_k = f(\Delta v_k),
\end{equation}
obtained by composing Eq.~(\ref{eq:massrec}) with the mission-specific score of the required wet mass. Since both the recursion and the score are monotone, $J_k$ increases with every leg increment, so that reducing a leg cost can never worsen the objective; the trade-off among the spacecraft, in contrast, is resolved by the outer search. The objective of the whole mission is
\begin{equation}
\label{eq:cost}
J = \sum_{k=1}^{N_{\rm sc}} J_k .
\end{equation}
The assignment of the targets to the spacecraft, together with the visiting orders and the rendezvous epochs, is handled by the evolutionary search framework that hosts the evaluator \cite{huang2026binpacking}; the contribution of this paper is the evaluator $F$ of Eq.~(\ref{eq:model}).

\subsection{Scope and Benchmark Instance}
\label{sec:scope}

The formulation above is instance-independent, and so is the model constructed for it: Eq.~(\ref{eq:model}) requires only (i) a pairwise leg-cost function for the transfer model of the mission and (ii) a sequence-level score that is monotone in the leg costs. It does not rely on the functional form of the score, on the perturbation model, or on the distribution of the targets. For a different mission, the same network architecture is retained and only the labeled dataset of Sec.~\ref{sec:data} is regenerated, by a procedure that is itself instance-independent.

For the numerical study reported below, one representative and demanding instance is used, namely the low-Earth-orbit multi-debris removal problem of the 9th Global Trajectory Optimization Competition (GTOC9) \cite{izzo2018kessler,jpl2018gtoc9}, in which 123 debris objects are to be removed by a fleet of spacecraft within a common mission window under $J_2$-perturbed impulsive transfers. The training labels, the evaluation samples, and the search results of this paper all refer to this instance, and its complete specification --- the orbital elements of the targets, the transfer model, the constraint values, and the scoring rule with its constants --- is given in Sec.~\ref{sec:instance}. This instance is chosen because it combines a large target set, a nonlinear mission score, and published reference solutions that allow a quantitative comparison; the method is not specific to it.

\section{Two-Layer Permutation-Invariant Cost Evaluation Network}
\label{sec:network}

\subsection{Overall Architecture}

\begin{figure}[htbp]
  \centering
  \includegraphics[width=\linewidth]{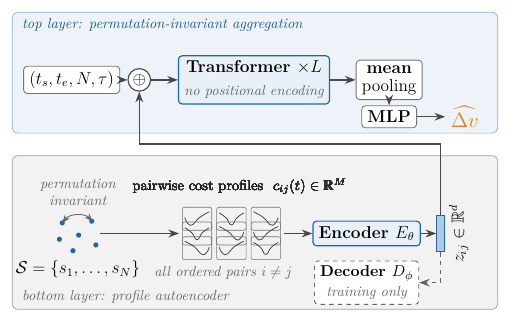}
  \caption{Overall architecture of the two-layer cost evaluation network. The bottom layer compresses the pairwise cost surface $c_{ij}(t,\Delta t)$ into a latent feature $\bm{z}_{ij}$; the top layer aggregates the latent features of all target pairs together with the orbit features and the window parameters by a permutation-invariant transformer encoder and regresses the total sequence cost.}
  \label{fig:arch}
\end{figure}

The design of the proposed model is motivated by an observed asymmetry of learnability: the cost of a single transfer between two targets has been shown to be learnable by deep networks with high accuracy \cite{hennes2016fast,zhu2019fast,zhang2026pretrained}, whereas the total cost of a sequence is difficult to learn directly because it involves the unknown optimal permutation and time assignment. The proposed model therefore adopts a two-layer structure, as illustrated in Fig.~\ref{fig:arch}. First, for each ordered target pair $(i,j)$, the pairwise transfer cost profile, i.e., the approximate minimum transfer cost as a function of the departure epoch within the window, is compressed into a latent feature vector by an autoencoder (denoted by the bottom layer). Then, the latent features of all target pairs in the set, together with the window parameters, are aggregated by a permutation-invariant attention network to regress the total sequence cost (denoted by the top layer). The overall model is expressed as
\begin{equation}
\label{eq:overall}
\widehat{\Delta v} = \mathcal{A}\!\left( \left\{ \bm{f}_i,\ \left\{ E\!\left(\bm{c}_{ij}\right) \right\}_{j \neq i} \right\}_{i \in \mathcal{S}},\ t_s,\ t_e \right),
\end{equation}
where $\bm{f}_i$ denotes the orbit feature of target $i$, $\bm{c}_{ij}$ denotes the pairwise cost surface between targets $i$ and $j$, $E(\cdot)$ denotes the encoder of the bottom layer, and $\mathcal{A}(\cdot)$ denotes the top-layer aggregation network. The pairwise latent features enter the aggregation only through permutation-invariant pooling over $j$ at each target $i$, so that no ordering information is introduced anywhere in the model.

\subsection{Bottom Layer: Compression of the Pairwise Cost Surface}
\label{sec:bottom}

\begin{figure}[htbp]
  \centering
  \includegraphics[width=\linewidth]{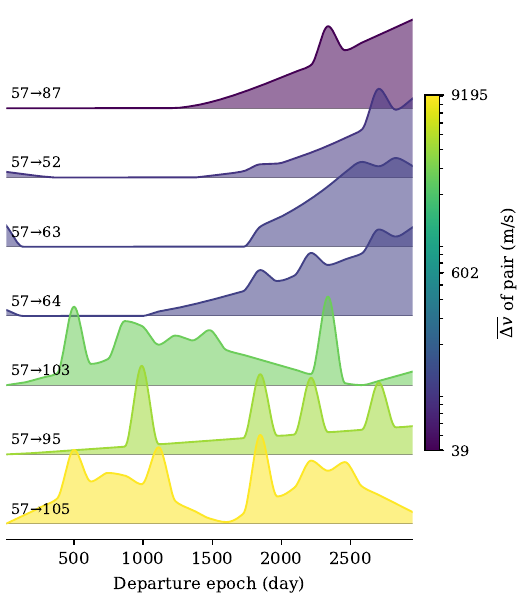}
  \caption{Examples of the pairwise cost profiles $c_{ij}(t)$ for one debris object of the benchmark target set (target 57 in the zero-based numbering of the 123 targets) and several of its partners $j$ from the same set. The profiles exhibit a prominent valley whose position and width are pair-specific, so that the whole function, rather than a point estimate, has to be carried into the cost evaluation.}
  \label{fig:profiles}
\end{figure}

For an ordered target pair $(i,j)$, the quantity to be represented is the pairwise cost surface $c_{ij}(t,\Delta t)$, i.e., the minimum transfer cost as a function of the departure epoch $t$ and the transfer duration $\Delta t$, produced by the semi-analytical single-leg approximation of \cite{huang2020fastestimation}. The accuracy of this approximation is sufficient for the present purpose: it was reported to reproduce the reference transfers of the GTOC9 scenario with a mean relative error of about 3\% while requiring less than a millisecond per evaluation \cite{huang2020fastestimation}. It is therefore adopted both as the surface from which the bottom-layer training samples are drawn and as the leg-cost function in the sequence-level optimization used for labeling (Sec.~\ref{sec:data}). The role of the bottom layer is to compress the information of this surface rather than to reproduce it accurately. For the target set of the benchmark instance, the surface varies strongly with both arguments: the low-cost windows occupy only a small part of the $t$--$\Delta t$ plane, and their locations differ from pair to pair, as illustrated in Fig.~\ref{fig:profiles} for one of these debris objects and several of its potential partners.

The bottom layer does not attempt to reproduce this surface accurately. Its role is \emph{information compression}: to encode the surface into a compact latent feature that preserves the trend information required by the top layer, namely the positions and widths of the low-cost valleys and the levels of the high-cost peaks, which together determine how the total sequence cost responds to the time assignment. To this end, the surface is sampled on a grid of $10 \times 8$ points, with normalized departure epochs $\bar{t} = 0, 0.1, \dots, 0.9$ and transfer durations $\Delta t = 1, 3, 5, 8, 12, 16, 20, 25$ d, and the bottom layer is trained as an autoencoder whose decoder is a coordinate network rather than a vector reconstructor:

\begin{equation}
\label{eq:ae}
\bm{z}_{ij} = E_\theta(\bm{x}_{ij}), \qquad
\hat{c}_{ij}(\bar{t}, \Delta t) = D_\phi\!\left(\bm{z}_{ij}, \bar{t}, \Delta t\right),
\end{equation}

where $\bm{x}_{ij}$ is the input feature of the pair and $\bm{z}_{ij} \in \mathbb{R}^{32}$ is the latent feature. The decoder concatenates the latent feature with the query coordinates and maps them to a cost value through a multilayer perceptron (256--256--128--1, ReLU); therefore $\bm{z}_{ij}$ must encode the whole surface as a continuous function of $(\bar{t}, \Delta t)$, instead of memorizing sampled values. The encoder is a multilayer perceptron (128--128--32, ReLU) that takes a 27-dimensional physics feature of the pair as input, composed of the orbit elements of the two targets and the relative phase of the pair at several epochs. The use of such physics features, rather than the sampled surface itself, is essential for the application: at search time the latent feature of any pair is produced by a single forward pass, without sweeping the $t$--$\Delta t$ plane.

The network parameters are trained by minimizing the relative reconstruction loss over all queries,
\begin{equation}
\label{eq:aeloss}
\mathcal{L}_{\rm AE} = \frac{1}{N_{\rm q}} \sum_{l=1}^{N_{\rm q}}
\left( \frac{\hat{c}_{l} - c_{l}}{c_{l}} \right)^2 ,
\end{equation}
with the Adam optimizer ($\mathrm{lr} = 10^{-3}$, batch size 512, 60 epochs) on $3.00 \times 10^{5}$ pair--window samples, each carrying the $10 \times 8$ query grid, i.e., about $2.4 \times 10^{7}$ labeled queries in total.

\subsection{Top Layer: Permutation-Invariant Aggregation Network}
\label{sec:top}

For a target set $\mathcal{S}$ with $N$ targets, the top layer builds one token per target rather than one token per pair, so that the token count grows linearly with the set size. The token of target $i$ combines its own orbit feature with the aggregated information of all pairwise latent features,
\begin{equation}
\label{eq:token}
\bm{h}_i^{(0)} = \varphi_n\!\left(\left[\bm{f}_i,\ \bar{t}_s,\ \bar{T}\right]\right)
+ \varphi_e\!\left(\left[\underset{j \neq i}{\mathrm{mean}}\ \bm{z}_{ij},\ \underset{j \neq i}{\max}\ \bm{z}_{ij}\right]\right),
\end{equation}
where $\varphi_n$ and $\varphi_e$ are linear projections to the model dimension $d = 64$, $\bm{f}_i \in \mathbb{R}^{9}$ is the orbit feature of target $i$ at the departure epoch $t_s$, and $\bar{t}_s = t_s / T_{\rm ref}$ and $\bar{T} = (t_e - t_s)/T_{\rm ref}$ are the normalized departure epoch and window length with $T_{\rm ref} = 650$ d. Both pooling operations over the partners $j$ are permutation-invariant, and the max term is retained deliberately: although the mean term describes the overall level of the pairwise costs, the cost of a sequence is often governed by a single dominant leg, whose information would be diluted by mean pooling alone. The tokens are then updated by $L = 2$ layers of multi-head self-attention with four heads, without any positional encoding,
\begin{equation}
\label{eq:mha}
\bm{H}^{(l+1)} = \mathrm{MHA}\!\left( \bm{H}^{(l)} W_Q,\; \bm{H}^{(l)} W_K,\; \bm{H}^{(l)} W_V \right) + \bm{H}^{(l)},
\end{equation}
followed by a feed-forward network with residual connections. Since no positional encoding is introduced and the self-attention is permutation-equivariant, the token representations are independent of the input order of the targets. The tokens are finally aggregated by masked mean and max pooling,
\begin{equation}
\label{eq:pool}
\bm{h} = \left[ \frac{1}{N} \sum_{i=1}^{N} \bm{h}_i^{(L)},\ \max_{i} \bm{h}_i^{(L)} \right],
\end{equation}
and the total sequence cost is regressed by a feed-forward head,
\begin{equation}
\label{eq:head}
\widehat{\Delta v} = \mathrm{MLP}\!\left(\left[\bm{h},\ \bar{t}_s,\ \bar{T}\right]\right),
\end{equation}
with a multilayer perceptron of size $(2d+2)$--128--64--1. Consequently, the overall mapping in Eq.~(\ref{eq:overall}) is invariant to any permutation of the input set, which is the desired property for order-undetermined evaluation. The set-learning foundation of this design can be found in \cite{zaheer2017deep,lee2019set}. It is worth emphasizing that, different from the attention encoders in the learned sequencing policies \cite{tomanek2025pointer,rivera2025neural}, where the set embedding serves as the context of an autoregressive decoder that constructs the sequence step by step, the aggregation here directly regresses a scalar cost and no ordering is predicted anywhere in the model.

The network is trained by minimizing the relative error loss
\begin{equation}
\label{eq:toploss}
\mathcal{L} = \frac{1}{N_{\rm seq}} \sum_{n=1}^{N_{\rm seq}} \left( \frac{\widehat{\Delta v}_n - \Delta v_n^*}{\Delta v_n^*} \right)^2 ,
\end{equation}
where $\Delta v_n^*$ is the label obtained by precise optimization (see Section~\ref{sec:data}). For reference, a variant trained with the mean-squared-error loss gives a similar accuracy and is not used further.

\subsection{Training Dataset Generation}
\label{sec:data}

\begin{figure*}[htbp]
  \centering
  \includegraphics[width=\textwidth]{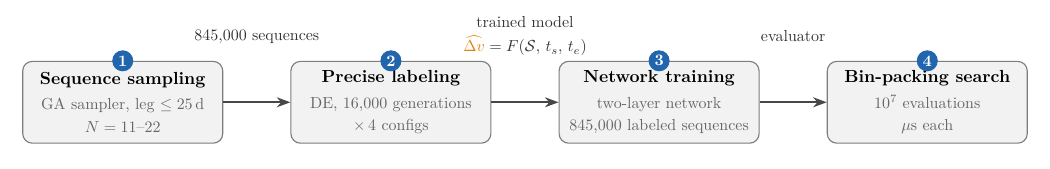}
  \caption{Generation of the training data and its use in the global search. The pairwise cost surface of every target pair is distilled by the bottom layer, the sequence cost is labeled by precise optimization, the top layer is trained on the labeled sequences, and the trained model then serves as the sequence evaluator of the evolutionary search.}
  \label{fig:pipeline}
\end{figure*}

\begin{figure*}[htbp]
  \centering
  \includegraphics[width=\textwidth]{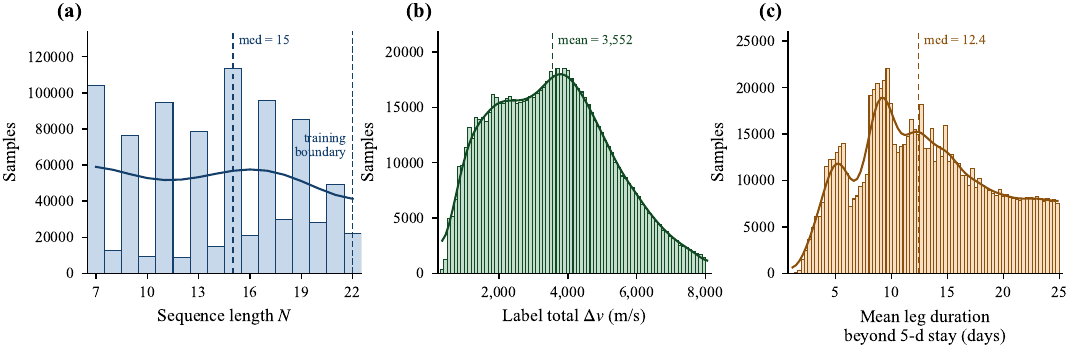}
  \caption{Statistics of the training dataset: (a) sequence length $N$; (b) labeled total velocity increment; (c) mean transfer duration per leg, excluding the 5-d stay.}
  \label{fig:data}
\end{figure*}

\begin{figure}[htbp]
  \centering
  \includegraphics[width=\linewidth]{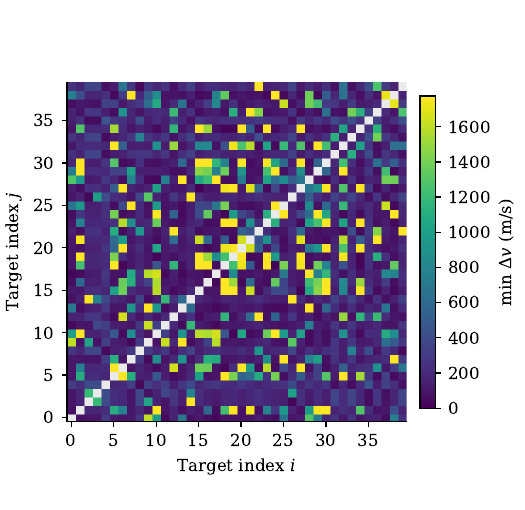}
  \caption{Minimum-over-epoch velocity increment of the target pairs (targets 0--39). The strong pair-to-pair variation is the reason why the pairwise information has to be carried into the sequence-level regression.}
  \label{fig:matrix}
\end{figure}

The training samples are generated by a two-step procedure, as illustrated in Fig.~\ref{fig:pipeline}. First, candidate sequences are sampled by a genetic algorithm: the sequence length is allowed to vary between 7 and 22 targets, each transfer duration is constrained not to exceed 25 d, and the remaining degrees of freedom (the departure epoch of the sequence and the internal rendezvous epochs) are generated randomly within the mission window of the benchmark instance specified in Sec.~\ref{sec:instance}; the leg costs are evaluated with the single-leg approximation of Sec.~\ref{sec:bottom}. Second, for each sampled sequence, the precise optimal velocity increment $\Delta v^*$ in Eq.~(\ref{eq:task}) is evaluated by differential evolution (DE) with the RandToBest1Exp strategy ($F = 0.9$, $CR = 0.8$, 16000 generations). Four configurations are run for every sequence --- a fixed departure epoch with a population of 53, the same with a population of 61, a departure epoch shifted forward by 30 d, and a departure epoch sliding within $[t_{s0},\, t_{s0}+80]$ d --- and the best result of the four is recorded as the label. A total of about 845{,}000 labeled sequences are collected, and their statistics are reported in Fig.~\ref{fig:data}: the sequence length ranges from 7 to 22 with a median of 15, the labeled total velocity increment concentrates between 2 and 4 km/s with a mean of 3.55 km/s, and the mean transfer duration per leg decreases from about 12.4 d as the sequence length increases, since the mission window is shared by more targets. The pair-to-pair variability of the quantity that the bottom layer has to represent is illustrated in Fig.~\ref{fig:matrix} for the first forty targets.

\section{Training Results and Application to GTOC9}
\label{sec:results}

\subsection{Benchmark Instance and Problem Specification}
\label{sec:instance}

The instance used in the numerical study is the multi-debris removal problem of GTOC9 \cite{izzo2018kessler,jpl2018gtoc9}, whose generic formulation was given in Sec.~\ref{sec:problem}. The orbital elements of the 123 debris objects are those distributed with the problem statement \cite{izzo2018kessler}, and the assumptions and constraint values of the instance are as follows. (1) All targets move on circular or slightly elliptical low Earth orbits; the $J_2$ term is the dominant perturbation at this altitude and is retained, while the higher-order gravitational and atmospheric terms are neglected \cite{huang2022global}. (2) All transfers are impulsive, and the spacecraft stays at each target for a fixed duration $\tau_{\rm stay} = 5$ d before the next transfer. (3) The departure epochs of the spacecraft lie within the common mission window of 2947 d, and the time gap between two successive missions is no less than 40 d. (4) Each of the 123 debris objects is removed exactly once, and a deorbit kit of $m_{\rm pl} = 30$ kg is released at every visited target.

The leg cost $\Delta v(i,j,t,t')$ of Sec.~\ref{sec:seqcost} is the minimum velocity increment of the $J_2$-perturbed multi-impulse rendezvous transfer between two targets with the prescribed stay, evaluated by the semi-analytical approximation of \cite{huang2020fastestimation}; with this model, Eq.~(\ref{eq:task}) defines the quantity to be estimated. The mass recursion of Eq.~(\ref{eq:massrec}) is evaluated with $I_{\rm sp} = 340$ s, $g_0 = 9.80665$ m/s$^2$, and $m_{\rm dry} = 2000$ kg. The mission score of the instance, i.e., the function $f(\cdot)$ of Eq.~(\ref{eq:costk}), is the quadratic function of the required wet mass,
\begin{equation}
\label{eq:costfunc}
J_k = 55 + 2\times10^{-6}\left(m_{0,k}-2000\right)^2 ,
\end{equation}
in units of MEUR, where $m_{0,k}$ is the wet mass at the beginning of the $k$-th sequence obtained from Eq.~(\ref{eq:massrec}) and the constant term accounts for the fixed cost associated with each spacecraft.

\subsection{Bottom-Layer Training Results}

\begin{figure}[htbp]
  \centering
  \includegraphics[width=\linewidth]{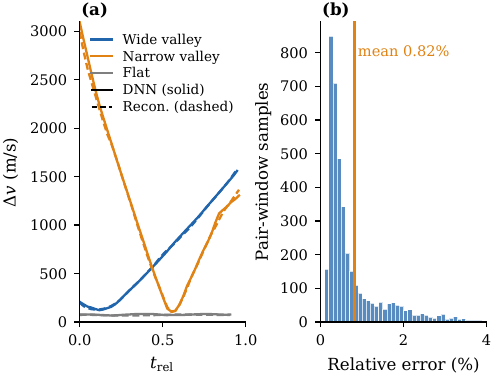}
  \caption{Reconstruction results of the bottom layer: (a) reference and reconstructed profiles in units of the transfer cost, for a wide-valley, a narrow-valley, and a nearly flat pair; (b) distribution of the relative reconstruction error over 4000 held-out pair--window samples.}
  \label{fig:ae}
\end{figure}

The reconstruction results of the bottom layer are illustrated in Fig.~\ref{fig:ae}. Over 4000 held-out pair--window samples, the mean relative error of the reconstructed surface is 0.82\%, with a median of 0.45\% and a 90th percentile of 1.93\%. More important than these values is that the reconstructed surfaces reproduce the structure of the reference ones for all the typical shapes of the analytical model --- wide valleys, narrow valleys, and nearly flat profiles --- including the positions and the depths of the low-cost valleys, which indicates that the compression retains the trend information of the pairwise cost function.

\begin{figure}[htbp]
  \centering
  \includegraphics[width=\linewidth]{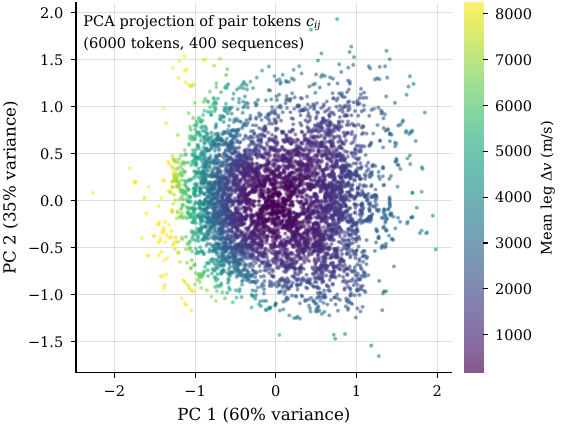}
  \caption{Two-dimensional embedding of the latent pairwise features $\bm{z}_{ij}$ of 6000 target pairs. The embedding is organized by the shape of the underlying cost surface, indicating that the bottom layer compresses the pairwise cost function into a representation that the top layer can compare across pairs.}
  \label{fig:latent}
\end{figure}

The latent representation learned in this way is visualized in Fig.~\ref{fig:latent}, where the pairwise features are embedded in two dimensions: the pairs are organized according to the distribution of their cost surfaces rather than grouped arbitrarily, so that the top layer receives a low-dimensional representation in which pairs with similar cost behaviour are placed close to each other. It should be re-emphasized that the bottom layer is not required to be an accurate surrogate of the analytical model by itself; its task is to provide the top layer with an informative and compact description of the pairwise cost function, and the accuracy of the sequence-level regression reported in Section~\ref{sec:topresults} is the practical evidence that this description is sufficient.

\subsection{Top-Layer Training Results}
\label{sec:topresults}

\begin{figure}[htbp]
  \centering
  \includegraphics[width=\linewidth]{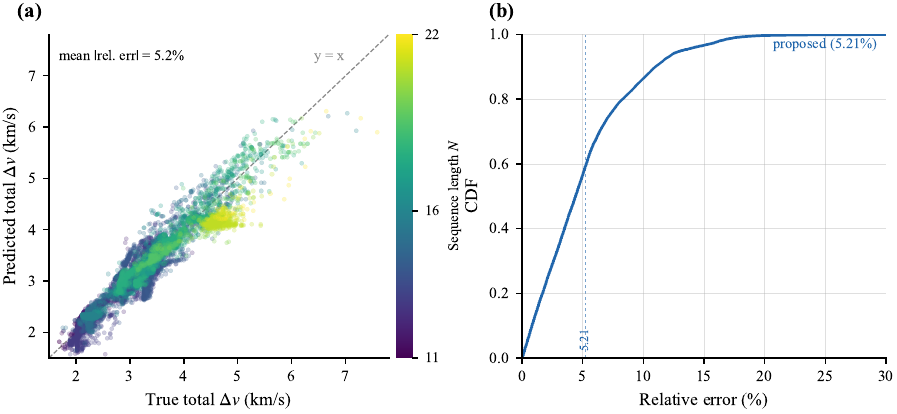}
  \caption{Accuracy of the top layer on the independent evaluation set of 18{,}415 sequences: (a) predicted against labeled total velocity increment, colored by sequence length; (b) cumulative distribution of the relative error.}
  \label{fig:accuracy}
\end{figure}

The trained aggregation network is evaluated on an independent set of 18{,}415 sequences that is disjoint from the training set; each sequence satisfies the mission-window and stay-time constraints, and its label is obtained by the same precise optimization as the training labels. The mean relative error of the estimated sequence cost is 5.2\%, with a median of 4.4\% and a 90th percentile of 11.0\%. The estimation of one sequence takes 0.94 ms on a single CPU core, whereas the precise DE verification of the same sequence (16000 generations, four configurations) requires 78.5 s on the same machine, which corresponds to a speedup of about $8\times10^{4}$, i.e., nearly five orders of magnitude. The agreement is illustrated in Fig.~\ref{fig:accuracy}: the scatter follows the diagonal over the whole range of the labeled velocity increment, and the error level is essentially independent of the sequence length, which is the expected consequence of a set-level regression that has no error-accumulation mechanism along the legs.

For reference, the error of a purely leg-wise aggregation of the same low-level analytical model was also examined. If every leg is assigned the departure epoch that minimizes its own cost on a 21-point grid while the true transfer durations of the optimized solution are retained, the mean relative error of the sequence estimate is 26.2\%; if the transfer duration of every leg is instead fixed at the tabulated maximum of 25 d, the mean relative error is 28.9\%. These two constructions neglect the joint time assignment and, in the second case, the actual leg durations; they are rough references that indicate the order of magnitude of the error produced by aggregating leg-wise estimates, and they are neither the accuracy of any published method nor the evidence of the advantage of the proposed model, whose accuracy is established by the comparison with the precisely optimized labels in Fig.~\ref{fig:accuracy}. It is also natural that such crude approximations would not be used directly in practice; they are reported here only to make the levels of accuracy explicit.

About 13.5\% of the evaluation sequences exhibit a relative error larger than 10\%, and only 0.29\% exceed 20\%. These sequences correspond to tight windows in which the optimal solution is forced to contain one leg whose cost is much larger than that of the others. Although the max pooling in Eq.~(\ref{eq:token}) preserves part of the information of the most expensive pair, the dominant leg still contributes only a small share of the pooled statistics, and a fixed-dimensional set representation cannot resolve the exact composition of a sequence \cite{wagstaff2019limitations}; consequently the model tends to underestimate the cost of such sequences. This limitation is compensated in the application by the final precise verification of the searched solutions, and it is also the motivation for the DE-in-the-loop stage described below.

\subsection{Search Framework and Settings}
\label{sec:gtoc9}

The proposed model is embedded into the bin-packing evolutionary framework of \cite{huang2026binpacking}, in which each target is treated as an item and each spacecraft as a bin, and the population is updated by customized mutation operators. The only modification is that the sequence cost evaluator is replaced by the proposed network. In the original framework, the sequence cost is evaluated by the order-free analytical approximation model \cite{huang2024orbit,huang2025cost}, whose mean relative error is about 10\%; the proposed network reduces this error to about 5\% while replacing the nonlinear program that the analytical evaluator solves for every candidate by a single forward pass, so that the evaluation becomes both more accurate and cheaper. The gain is twofold. First, since the evaluation of one sequence takes less than one millisecond, a single search campaign evaluates on the order of $10^7$ candidate sequences within hours on a multi-core workstation, which is unaffordable for optimization-based evaluators. The computational effort that this requires in the present implementation is also worth noting. In the present implementation, the evaluation of one candidate sequence costs 0.94 ms on a single core, so that one run of the evolutionary search, which performs hundreds to a few thousand generations, evaluates on the order of $10^5$ candidate sequences within a few minutes. The solutions reported below were obtained from tens to several hundred independent runs per scale, that is, from of the order of $10^7$ to $10^8$ evaluations in total, and each promising candidate retained by the network was subsequently verified by precise optimization at a cost of 78.5 s per sequence (four configurations of 16000 generations), i.e., about 12 min for a nine-sequence solution. A detailed comparison of the computational effort and of the reported costs of the published global search methods is given in Ref.~\cite{huang2026binpacking} and is not repeated here. Second, the reduced estimation error directly lowers the probability that a truly superior sequence is discarded because of an approximation error, which is essential for a search whose selection pressure acts entirely on the estimated costs. In each campaign, the initial solutions are generated randomly with nonuniform mission durations, and three mutation operators (move, swap, and time adjustment) are applied; each cost-increasing mutation is accepted with a probability of 30\%.

\subsection{Global Search Results}

Search campaigns are conducted with $M=7$, $8$, $9$, and $10$ spacecraft, respectively, and the direct outputs of the search, i.e., the solutions reported by the network before any precise re-optimization, are examined first. Table~\ref{tab:gtoc9} compares the obtained solutions with the best published results of the corresponding scales. For $M=7$, the network evaluates the best candidate at 705.3 MEUR, and precise trajectory optimization gives 800.4 MEUR, which improves the best published solution of 844 MEUR \cite{zhang2022twostage} by 5.2\%. For $M=10$, the direct output of 708.7 MEUR is verified to be 702.8 MEUR, which improves the 731 MEUR of the GTOC9 winning solution \cite{jpl2018gtoc9}; it is noted that the verified cost here is even lower than the estimate, since the verification re-optimizes the rendezvous epochs of every sequence. For $M=8$, the direct output of 660.3 MEUR is verified to be 713.5 MEUR, which is within 2.8\% of the best known 694 MEUR \cite{zhang2022twostage} and already better than the 730 MEUR obtained by the same bin-packing framework with the analytical evaluator \cite{huang2026binpacking}. For $M=9$, the best direct output obtained from random initializations is evaluated by the network at 659.4 MEUR and verified to be 677.3 MEUR, which is 4.7\% above the best published nine-spacecraft result of 647.1 MEUR \cite{zhang2024global}; as will be shown below, this scale requires either the refinement stage or a good initial solution to reach the published level.

This behavior deserves emphasis, because it contrasts with the way in which the best solutions of previous global search methods were obtained. In nearly all existing studies, the reported best solution is an outlier: a massive number of repeated campaigns first produce a large population of local solutions, and an excellent individual emerges from this population only with a small probability. The framework adopted in this paper inherits this search mechanism, yet the verified results show that high-quality solutions are now produced with a considerably higher probability. The verification of the search outputs (with the campaigns of all the workstations merged, since they share the same samplers and evaluators) gives a clear picture of the population that the search produces. At $M=7$, the verified costs span 800.4--1159.3 MEUR, of which three improve the best published 844 MEUR; the wide spread at this scale is the direct consequence of the strongest amplification of the estimation error, since the sequences of this scale are the longest and the per-spacecraft velocity increments the largest, and it is the reason why the outputs of this scale must always be ranked by precise verification. At $M=8$ the verified costs span 707.5--839.0 MEUR, six of which are better than the 730 MEUR obtained by the bin-packing framework with the analytical evaluator \cite{huang2026binpacking}. At $M=9$ they span 650.1--758.7 MEUR, and when the search is started from the best published nine-spacecraft solution the final result reaches 629.8 MEUR, which is better than both the 647.1 MEUR of the best published solution \cite{zhang2024global} and the 657.3 MEUR obtained in the authors' previous work with the analytical evaluator \cite{huang2022global}. At $M=10$, 24 verified outputs deviate from the self-reported ones by only $+2.3\%\pm1.7\%$ on average, the best of them reaching 702.8 MEUR although it ranked only twelfth by the self-reported cost. This rank mismatch is precisely the intended behavior of the two-stage design: the network serves as a coarse filter that retains a population of candidates containing the truly superior sequences, while the final ranking is always established by precise verification.

\begin{figure}[htbp]
  \centering
  \includegraphics[width=\linewidth]{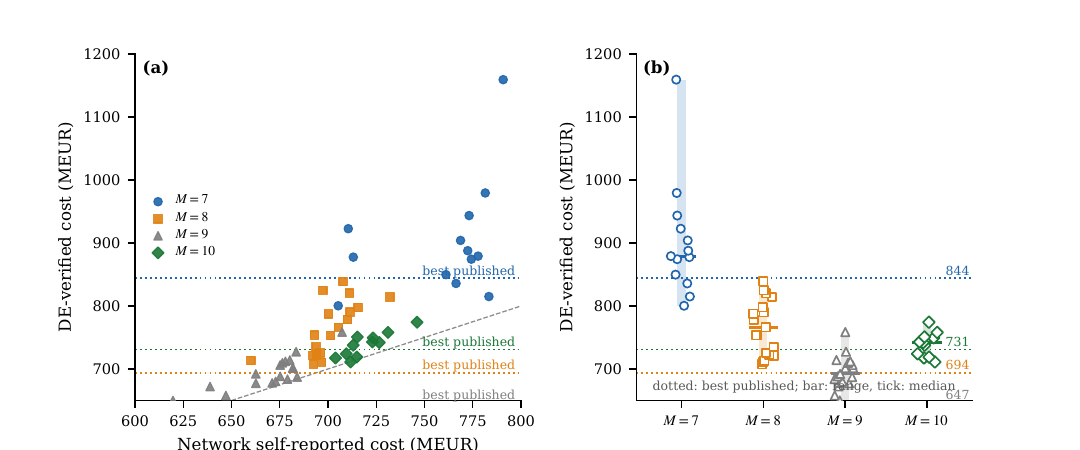}
  \caption{Precise verification of the search outputs, with the campaigns of all workstations merged: (a) self-reported cost against DE-verified cost for the four mission scales, with the best published values marked as dotted lines; (b) distribution of the verified costs of the direct outputs around the best published values.}
  \label{fig:verify}
\end{figure}

In other words, the proposed evaluator does not replace the precise evaluation; it makes it affordable to screen a very large number of candidates, so that the population submitted to verification contains solutions of consistently high quality. The verified costs of the direct outputs are summarized in Fig.~\ref{fig:verify}.

\subsection{Refinement of the Network Outputs}

Besides being evaluated directly, the solutions produced by the network are also a convenient starting point for a more accurate optimization. In the global search framework adopted here, a sequence modified by a mutation operator can be re-evaluated by precise optimization, so that the search progressively replaces the network estimates by precise costs; the bin-packing evolutionary framework and its in-loop mechanism are described in Ref.~\cite{huang2026binpacking} and are not repeated here, since the contribution of this paper is the sequence cost evaluator rather than the framework. The purpose of this subsection is to show that the network estimates are accurate enough to serve as effective initial solutions for the precise optimization.

Starting from the direct outputs, the refinement reaches 737.8 and 676.1 MEUR for $M=7$ and $M=8$, which improve the best published values of these scales by 12.6\% and 2.6\%; for $M=9$ the refinement of the direct output reaches 660.1 MEUR, and when the search is warm-started from the best published nine-spacecraft solution, the refinement improves that record to 629.8 MEUR, i.e., by 2.7\%. For $M=10$, the verified direct output of 702.8 MEUR already improves the winning solution and no refinement is reported; many candidates that the network evaluates at about 700 MEUR are verified in the range of 70x--71x MEUR, which indicates that the estimation error is of the least consequence at this scale.

\begin{figure*}[htbp]
  \centering
  \includegraphics[width=\textwidth]{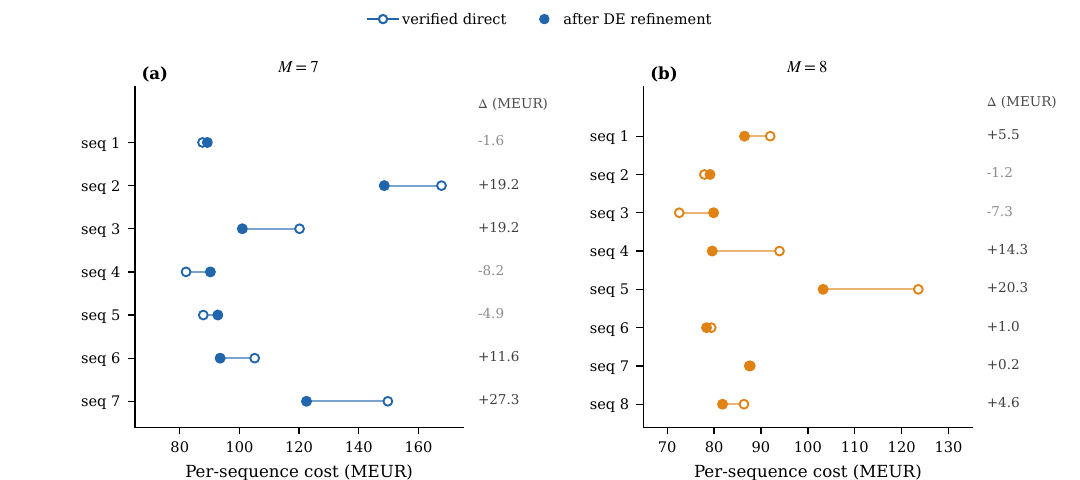}
  \caption{Per-sequence comparison of the direct output and the refined solution for the seven- and eight-spacecraft campaigns. The refinement not only adjusts the rendezvous epochs but also moves targets between neighbouring sequences, so that the improvement is concentrated in the sequences whose cost was initially the highest.}
  \label{fig:dumbbell}
\end{figure*}

Two observations follow from these results. First, the refinement changes the solutions in a structured way: the per-sequence comparison of Fig.~\ref{fig:dumbbell} shows that the rendezvous epochs are adjusted and targets are moved between neighbouring sequences, and that the largest improvement is concentrated in the sequences whose cost was initially the highest. Second, the number of precise optimizations required is modest, because the network screens the candidates in advance and only the sequences modified by the operators are re-evaluated. In this sense the network does not replace the precise optimization; it reduces the number of candidates that the latter has to examine, which is precisely the efficiency advantage that this paper targets.

\begin{table}[t]
  \centering
  \caption{Comparison of the GTOC9 solutions (all costs verified by precise trajectory optimization). The published row gives the best result reported for each scale, compiled from Refs.~\cite{zhang2022twostage,zhang2024global,huang2026binpacking,jpl2018gtoc9}; these methods are compared with one another in Ref.~\cite{huang2026binpacking} and are not re-implemented here, and the comparison in this paper is therefore made only against the best published value. $^\dagger$ denotes the result obtained by refining the best published nine-spacecraft solution.}
  \label{tab:gtoc9}
  \scriptsize
  \setlength{\tabcolsep}{3pt}
  \begin{tabular}{lcccc}
    \toprule
    Solution & $M=7$ & $M=8$ & $M=9$ & $M=10$ \\
    \midrule
    Best published & 844 & 694 & 647.1 & 731 \\
    \midrule
    This paper, direct output & 800.4 & 713.5 & 677.3 & 702.8 \\
    This paper, after refinement & \textbf{737.8} & \textbf{676.1} & 660.1 & --- \\
    This paper, refinement$^\dagger$ & --- & --- & \textbf{629.8} & --- \\
    \bottomrule
  \end{tabular}
\end{table}

\begin{table*}[t]
  \centering
  \caption{Per-sequence comparison of the network direct output and the DE-refined solution ($M=7$). For the direct output, $\Delta v$ is the network estimate and the verified values are obtained by precise optimization of each sequence (16000-generation DE, four configurations); for the refined solution, all values are precise-optimization results.}
  \label{tab:perseq7}
  \scriptsize
  \setlength{\tabcolsep}{3pt}
  \begin{tabular}{cclrrrp{5.6cm}}
    \toprule
    Seq. & $N$ & Window, day & $\Delta v^{\mathrm{net}}$, m/s & $\Delta v^{\mathrm{DE}}$, m/s & $J^{\mathrm{DE}}$, MEUR & Target assignment \\
    \midrule
    \multicolumn{7}{l}{\emph{$M=7$ (direct self-report 705.3)}} \\
    1 & 18 & 10--383 & 2792 & 3136 & 87.63 & 121, 84, 16, 118, 27, 20, 57, 23, 46, 87, 114, 38, 120, 103, 95, 50, 22, 15 \\
    2 & 21 & 423--815 & 4080 & 4698 & 167.71 & 0, 122, 2, 77, 104, 26, 68, 37, 75, 88, 63, 94, 106, 11, 21, 102, 18, 33, 67, 47, 70 \\
    3 & 16 & 855--1242 & 4006 & 4071 & 120.14 & 1, 40, 62, 74, 80, 99, 105, 32, 89, 86, 116, 117, 17, 59, 101, 98 \\
    4 & 17 & 1282--1611 & 2868 & 2982 & 82.11 & 5, 12, 78, 39, 61, 42, 43, 53, 56, 60, 58, 25, 71, 109, 92, 107, 111 \\
    5 & 16 & 1651--1910 & 3169 & 3204 & 87.89 & 83, 44, 9, 69, 90, 115, 10, 28, 65, 66, 79, 97, 113, 34, 110, 93 \\
    6 & 17 & 1950--2385 & 3695 & 3737 & 105.12 & 7, 14, 54, 36, 24, 41, 45, 52, 64, 49, 72, 8, 76, 108, 55, 119, 82 \\
    7 & 18 & 2425--2857 & 3672 & 4535 & 149.74 & 91, 73, 29, 31, 100, 48, 6, 51, 96, 112, 81, 35, 85, 4, 13, 3, 30, 19 \\
    \midrule
    \multicolumn{7}{l}{\emph{$M=7$ (refined, verified 737.80)}} \\
    1 & 17 & 10--399 & --- & 3230 & 89.25 & 105, 95, 57, 120, 103, 46, 22, 38, 15, 16, 87, 23, 121, 118, 20, 50, 27 \\
    2 & 21 & 439--803 & --- & 4415 & 148.54 & 0, 122, 2, 104, 26, 68, 37, 75, 88, 82, 45, 94, 106, 11, 21, 102, 18, 33, 67, 47, 70 \\
    3 & 17 & 835--1219 & --- & 3591 & 100.98 & 116, 117, 62, 80, 101, 86, 1, 40, 74, 17, 99, 114, 59, 98, 96, 32, 89 \\
    4 & 17 & 1259--1594 & --- & 3267 & 90.26 & 107, 61, 78, 60, 4, 58, 5, 109, 53, 111, 12, 92, 71, 56, 39, 42, 43 \\
    5 & 17 & 1634--1904 & --- & 3350 & 92.76 & 83, 44, 9, 49, 69, 90, 115, 10, 97, 28, 65, 66, 79, 113, 34, 110, 93 \\
    6 & 16 & 1944--2382 & --- & 3429 & 93.54 & 52, 63, 84, 36, 72, 14, 108, 41, 76, 8, 64, 24, 54, 7, 55, 119 \\
    7 & 18 & 2422--2856 & --- & 4054 & 122.47 & 19, 48, 51, 35, 30, 100, 73, 6, 81, 29, 31, 77, 112, 25, 91, 3, 13, 85 \\
    \midrule
    \multicolumn{3}{l}{Total} & 24282 & 26363 / 25336 & 800.35 / 737.80 & \\
    \bottomrule
  \end{tabular}
\end{table*}

\begin{table*}[t]
  \centering
  \caption{Per-sequence comparison of the network direct output and the DE-refined solution ($M=8$). For the direct output, $\Delta v$ is the network estimate and the verified values are obtained by precise optimization of each sequence (16000-generation DE, four configurations); for the refined solution, all values are precise-optimization results.}
  \label{tab:perseq8}
  \scriptsize
  \setlength{\tabcolsep}{3pt}
  \begin{tabular}{cclrrrp{5.6cm}}
    \toprule
    Seq. & $N$ & Window, day & $\Delta v^{\mathrm{net}}$, m/s & $\Delta v^{\mathrm{DE}}$, m/s & $J^{\mathrm{DE}}$, MEUR & Target assignment \\
    \midrule
    \multicolumn{7}{l}{\emph{$M=8$ (direct self-report 660.0)}} \\
    1 & 19 & 10--458 & 3100 & 3252 & 91.97 & 15, 22, 35, 38, 46, 120, 57, 14, 62, 96, 81, 87, 103, 105, 114, 95, 16, 117, 97 \\
    2 & 17 & 498--833 & 2641 & 2795 & 77.93 & 122, 70, 18, 39, 37, 63, 11, 26, 2, 75, 47, 67, 88, 104, 85, 21, 6 \\
    3 & 12 & 873--1058 & 2742 & 2658 & 72.57 & 42, 31, 50, 53, 61, 29, 107, 111, 5, 25, 56, 78 \\
    4 & 15 & 1098--1375 & 3046 & 3403 & 93.94 & 77, 9, 68, 72, 28, 94, 33, 55, 69, 90, 93, 52, 64, 3, 115 \\
    5 & 17 & 1415--1833 & 3534 & 4063 & 123.56 & 13, 17, 116, 4, 51, 59, 76, 80, 98, 91, 102, 99, 121, 0, 48, 101, 112 \\
    6 & 13 & 1873--2224 & 2562 & 2976 & 79.40 & 10, 19, 44, 65, 73, 34, 100, 30, 83, 110, 79, 86, 92 \\
    7 & 14 & 2264--2537 & 2878 & 3270 & 87.73 & 32, 8, 49, 109, 36, 7, 118, 84, 113, 54, 74, 119, 24, 66 \\
    8 & 16 & 2577--2937 & 3205 & 3150 & 86.37 & 60, 1, 12, 27, 40, 58, 41, 45, 71, 82, 106, 89, 20, 108, 23, 43 \\
    \midrule
    \multicolumn{7}{l}{\emph{$M=8$ (refined, verified 676.07)}} \\
    1 & 18 & 10--466 & --- & 3103 & 86.49 & 57, 95, 62, 81, 105, 87, 22, 46, 114, 14, 97, 103, 35, 120, 96, 16, 38, 117 \\
    2 & 18 & 506--837 & --- & 2795 & 79.13 & 122, 70, 94, 18, 39, 37, 63, 0, 11, 26, 2, 75, 47, 88, 104, 85, 21, 6 \\
    3 & 13 & 876--1053 & --- & 3001 & 79.89 & 29, 107, 61, 53, 56, 79, 50, 42, 111, 31, 5, 25, 78 \\
    4 & 14 & 1093--1376 & --- & 2940 & 79.60 & 30, 9, 68, 72, 28, 33, 55, 69, 90, 93, 52, 64, 3, 115 \\
    5 & 18 & 1416--1852 & --- & 3623 & 103.25 & 13, 17, 116, 15, 4, 51, 59, 76, 67, 80, 98, 91, 99, 121, 20, 48, 101, 112 \\
    6 & 13 & 1884--2217 & --- & 2923 & 78.39 & 77, 19, 44, 83, 73, 10, 100, 65, 34, 110, 86, 102, 92 \\
    7 & 14 & 2257--2519 & --- & 3260 & 87.52 & 32, 8, 49, 109, 36, 7, 118, 84, 113, 54, 74, 119, 24, 66 \\
    8 & 15 & 2555--2937 & --- & 3004 & 81.80 & 60, 1, 12, 27, 40, 58, 41, 45, 71, 82, 106, 89, 108, 23, 43 \\
    \midrule
    \multicolumn{3}{l}{Total} & 23709 & 25567 / 24648 & 713.47 / 676.07 & \\
    \bottomrule
  \end{tabular}
\end{table*}

\begin{table*}[t]
  \centering
  \caption{Per-sequence data of the $M=9$ solution obtained by refining the published nine-spacecraft solution \cite{zhang2024global}; the costs of the nine sequences sum to 629.8 MEUR (Table~\ref{tab:gtoc9}). All values are precise-optimization results obtained with the arrival epochs of the refined solution.}
  \label{tab:perseq9}
  \scriptsize
  \setlength{\tabcolsep}{3pt}
  \begin{tabular}{cclrrp{6.8cm}}
    \toprule
    Seq. & $N$ & Window, day & $\Delta v^{\mathrm{DE}}$, m/s & $J^{\mathrm{DE}}$, MEUR & Target assignment \\
    \midrule
    1 & 16 & 5--362 & 2976 & 81.82 & 114, 120, 27, 38, 20, 57, 118, 117, 95, 55, 113, 79, 25, 84, 16, 87 \\
    2 & 15 & 402--703 & 2205 & 67.64 & 112, 58, 76, 29, 51, 72, 73, 66, 10, 107, 61, 28, 64, 3, 52 \\
    3 & 14 & 739--973 & 2833 & 77.33 & 102, 33, 18, 75, 26, 11, 63, 34, 88, 45, 82, 41, 7, 70 \\
    4 & 13 & 1013--1250 & 2184 & 66.55 & 48, 83, 32, 105, 46, 89, 74, 22, 119, 104, 24, 108, 37 \\
    5 & 13 & 1290--1548 & 2019 & 64.49 & 106, 68, 100, 93, 30, 9, 69, 90, 77, 94, 19, 21, 49 \\
    6 & 16 & 1584--1909 & 2138 & 67.02 & 36, 1, 40, 62, 99, 54, 122, 85, 59, 98, 35, 67, 116, 8, 15, 121 \\
    7 & 12 & 1949--2198 & 2842 & 76.28 & 101, 43, 53, 5, 71, 13, 60, 12, 39, 23, 80, 47 \\
    8 & 12 & 2234--2563 & 2019 & 64.27 & 97, 50, 31, 96, 115, 81, 6, 65, 91, 2, 92, 110 \\
    9 & 12 & 2656--2941 & 2025 & 64.40 & 44, 56, 111, 103, 17, 78, 0, 109, 86, 14, 42, 4 \\
    \midrule
    \multicolumn{3}{l}{Total} & 21242 & 629.80 & \\
    \bottomrule
  \end{tabular}
\end{table*}

\begin{table*}[t]
  \centering
  \caption{Per-sequence data of the network direct output ($M=10$). $\Delta v$ is the network estimate; the verified values are obtained by precise optimization of each sequence (40000-generation DE, two configurations).}
  \label{tab:perseq10}
  \scriptsize
  \setlength{\tabcolsep}{3pt}
  \begin{tabular}{cclrrrp{5.6cm}}
    \toprule
    Seq. & $N$ & Window, day & $\Delta v^{\mathrm{net}}$, m/s & $\Delta v^{\mathrm{DE}}$, m/s & $J^{\mathrm{DE}}$, MEUR & Target assignment \\
    \midrule
    \multicolumn{7}{l}{\emph{$M=10$ (direct self-report 708.7)}} \\
    1 & 14 & 10--217 & 2116 & 2258 & 67.90 & 2, 56, 60, 29, 102, 28, 66, 111, 42, 90, 36, 86, 72, 110 \\
    2 & 15 & 257--524 & 2744 & 2767 & 76.65 & 55, 113, 79, 20, 118, 50, 27, 84, 83, 14, 121, 23, 97, 16, 103 \\
    3 & 14 & 564--855 & 2826 & 2654 & 73.94 & 30, 19, 43, 100, 71, 31, 65, 52, 69, 10, 73, 91, 12, 93 \\
    4 & 12 & 895--1145 & 2998 & 2762 & 74.75 & 104, 6, 18, 37, 88, 34, 70, 33, 68, 106, 77, 115 \\
    5 & 10 & 1185--1347 & 2872 & 3016 & 78.57 & 39, 8, 53, 4, 58, 92, 78, 5, 61, 107 \\
    6 & 16 & 1387--1673 & 2525 & 2464 & 71.83 & 22, 75, 81, 82, 41, 87, 105, 96, 46, 45, 3, 63, 114, 32, 108, 117 \\
    7 & 15 & 1713--2057 & 2308 & 2245 & 68.17 & 1, 40, 62, 89, 0, 99, 112, 15, 59, 98, 116, 122, 85, 17, 80 \\
    8 & 9 & 2097--2208 & 1707 & 1775 & 61.22 & 48, 25, 94, 120, 38, 51, 26, 101, 35 \\
    9 & 11 & 2248--2454 & 2234 & 2361 & 67.93 & 64, 95, 76, 54, 109, 49, 24, 7, 74, 119, 9 \\
    10 & 7 & 2494--2620 & 1910 & 1926 & 61.84 & 11, 47, 21, 13, 44, 67, 57 \\
    \midrule
    \multicolumn{3}{l}{Total} & 23541 & 24226 & 702.79 & \\
    \bottomrule
  \end{tabular}
\end{table*}

\subsection{Effect of the Mission Scale}
\label{sec:scale}

It should be noted that all the costs reported above are the verified results of precise trajectory optimization rather than the raw network outputs, and that the gap between them depends strongly on the mission scale. Two mechanisms contribute to this gap. The first is the well-known selection bias of surrogate-assisted search \cite{smith2006optimizer,jin2002framework}: the selection pressure preferentially retains the sequences whose costs are underestimated by the model. The second is the sensitivity of the cost metric itself: the cost is an exponential-type function of the velocity increments through the mass recursion, so that a given relative $\Delta v$ error is amplified for sequences with large $\Delta v$. In the $\Delta v$ metric the agreement is considerably better; for example, the total-$\Delta v$ errors of the two direct outputs of Table~\ref{tab:perseq7} and Table~\ref{tab:perseq8} are $+8.6\%$ and $+7.8\%$, and most individual sequences agree within a few percent, whereas the cost gap is dominated by one or two large-$\Delta v$ sequences (e.g., the last sequence of $M=7$, whose $\Delta v$ error of $+23.5\%$ translates into a cost error of $+40.7\%$).

\begin{figure}[htbp]
  \centering
  \includegraphics[width=\linewidth]{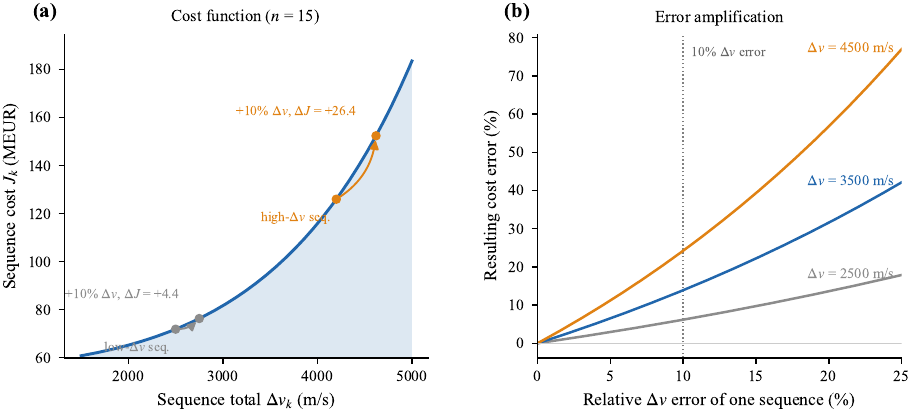}
  \caption{Amplification of the velocity-increment error into the cost metric: (a) cost of a 15-target sequence as a function of its total velocity increment; (b) resulting cost error for a given relative error of the estimated velocity increment, at three levels of $\Delta v$.}
  \label{fig:ampl}
\end{figure}

The same trend is observed systematically across the mission scales: the per-sequence $\Delta v$ mean absolute error of the verified direct outputs remains within 6.1\%--9.6\% for $M=7$ to $10$, and the solution-level total-$\Delta v$ errors are $+9.5\%$, $+6.8\%$, $+3.9\%$, and $+4.7\%$, which shows that the estimation quality in the $\Delta v$ metric is essentially uniform over the mission scale. The cost-metric gap, in contrast, decreases from $+17.1\%$ ($M=7$) to $+9.3\%$ ($M=8$), $+2.6\%$ ($M=9$), and $+2.6\%$ ($M=10$), which is a property of the convex cost mapping rather than of the evaluator: the local elasticity of the verified cost with respect to the total $\Delta v$, empirically 1.41, 1.03, 0.66, and 0.56 for $M=7$ to $10$, weakens as the per-sequence $\Delta v$ decreases when more spacecraft share the same targets. Fig.~\ref{fig:ampl} illustrates this mechanism: a given relative error of the velocity increment is translated into a much larger relative error of the cost for large-$\Delta v$ sequences, so that the accuracy of the evaluation has to be judged in the $\Delta v$ metric, in which the model is uniformly accurate across the scales, while the cost metric is the quantity that must be verified.

It should be emphasized that the amplification is a property of the mission metric rather than of the estimator. In the velocity-increment metric the deviation of the same solutions is modest, as detailed in Table~\ref{tab:gap}: for the four headline solutions the total-$\Delta v$ differences are $+8.6\%$, $+7.8\%$, $+4.4\%$, and $+2.9\%$ for $M=7$ to $10$, while their cost differences are $+13.5\%$, $+8.1\%$, $+2.7\%$, and $-0.8\%$. Over all the verified outputs of each scale, the mean $\Delta v$ gaps are those listed above, whereas the mean cost gaps reach $+17.1\%$, $+9.3\%$, $+2.6\%$, and $+2.6\%$. The cost metric defined by the GTOC9 rules is a quadratic function of the wet mass, $J_k = 55 + 2\times10^{-6}(m_{0,k}-2000)^2$, and the wet mass grows exponentially with the velocity increments through the recursion of Eq.~(\ref{eq:massrec}); the sensitivity of the metric to a given relative error therefore increases rapidly with the level of $\Delta v$, and the high-$\Delta v$ sequences of the small-fleet solutions dominate their cost. Accordingly, the two gaps are of the same order at $M=8$ to $M=10$, where the per-spacecraft $\Delta v$ is low, whereas the metric effect becomes pronounced at $M=7$, where a $\Delta v$ gap below 10\% is translated into a cost gap above 17\%.

\begin{table*}[htbp]
  \centering
  \caption{Estimation gap of the search outputs in the velocity-increment and cost metrics. The two numerical columns in the middle refer to the direct outputs of the search, i.e., the solutions reported by the network before any precise re-optimization (Tables~\ref{tab:perseq7}--\ref{tab:perseq10}; for $M=9$, the direct output obtained from random initializations, whereas the refined nine-spacecraft solution of Table~\ref{tab:perseq9} is not a network output and is not listed here); the last two columns are the mean gaps over all the verified outputs of each scale, with the campaigns of all workstations merged.}
  \label{tab:gap}
  \small
  \begin{tabular}{lcccccc}
    \toprule
    $M$ & $\Delta v^{\rm net}$, m/s & $\Delta v^{\rm DE}$, m/s & $\Delta v$ gap & $J^{\rm net}$, MEUR & $J^{\rm DE}$, MEUR & Cost gap (headline / mean) \\
    \midrule
    7  & 24{,}282 & 26{,}363 & $+8.6\%$ & 705.3 & 800.4 & $+13.5\%$ / $+17.1\%$ \\
    8  & 23{,}709 & 25{,}567 & $+7.8\%$ & 660.0 & 713.5 & $+8.1\%$ / $+9.3\%$ \\
    9  & 23{,}115 & 24{,}137 & $+4.4\%$ & 659.4 & 677.3 & $+2.7\%$ / $+2.6\%$ \\
    10 & 23{,}541 & 24{,}226 & $+2.9\%$ & 708.7 & 702.8 & $-0.8\%$ / $+2.6\%$ \\
    \bottomrule
  \end{tabular}
\end{table*}

This mechanism explains the different behavior of the four scales. Since the targets are shared by more sequences as $M$ increases, the velocity increment of an individual spacecraft decreases, and the same relative estimation error is translated into a smaller error of the mission cost, as quantified by the elasticities above. The consequence is visible in the results. At $M=10$, the cost is the least sensitive, and the direct outputs are already competitive without any precise re-optimization. At $M=9$, although the mean gap is small, an individual solution may still contain a sequence whose $\Delta v$ is underestimated by about 5\%, which corresponds to tens of MEUR in the cost of that solution (for example, 676.4 MEUR estimated against 709.9 MEUR verified); neither the ranking nor the record of this scale can therefore be established without precise evaluation, and this is why the from-scratch result of \mbox{$M=9$} remains above the published record. At $M=8$, the direct output is close to, but not yet better than, the published value, and the refined solution surpasses it clearly. At $M=7$, in contrast, the fewest spacecraft share the same targets and the per-spacecraft $\Delta v$ is the largest, so that the cost is the most sensitive to the estimation error of all scales; nevertheless, the best direct outputs of this scale still improve the published record, which is the most direct evidence of the advantage of the proposed evaluator for long target sequences: even though the network error is amplified by the convex cost mapping, the gain brought by the much faster and order-free evaluation of the same sequences outweighs this amplification. The sequence that dominates the gap of the $M=7$ direct output illustrates the amplification and its remedy: the network estimates 3672 m/s for this sequence whereas precise optimization gives 4535 m/s, i.e., a $\Delta v$ error of $+23.5\%$ and a cost error of $+40.7\%$; after the refinement, the same sequence is re-ordered and re-timed to 4054 m/s, and its cost is reduced from 149.7 to 122.5 MEUR, which constitutes the dominant part of the overall improvement of the $M=7$ solution. When the refined $M=7$ solution is evaluated by the network as a whole, the estimate is 749.0 MEUR against the verified 737.8 MEUR, i.e., a total-$\Delta v$ estimate of 25{,}668 m/s against 25{,}336 m/s (+1.3\%), so that the gap of the direct output essentially disappears: the direct output had been pushed by the search into a region where the network underestimates the cost, whereas the refined solution lies well inside the distribution represented by the training data.

\begin{figure}[htbp]
  \centering
  \includegraphics[width=\linewidth]{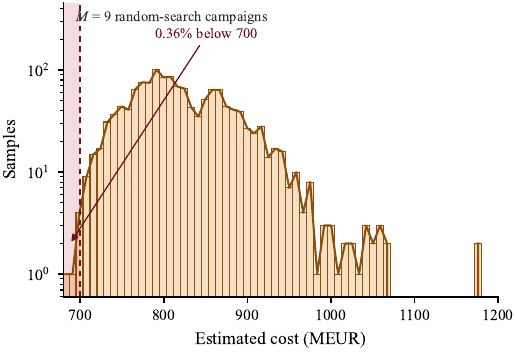}
  \caption{Distribution of the self-reported costs of 21{,}163 solutions obtained from completely random initializations (nine spacecraft), on a logarithmic scale. Solutions below 700 MEUR are rare tail events with a proportion of 0.40\%.}
  \label{fig:hist}
\end{figure}

In the pure random-search path, in which no prior knowledge of existing solutions is used, the best solution obtained from completely random initializations is verified to be 687.8 MEUR ($M=9$), which is still better than most of the competition submissions. The distribution of the estimated costs of the random-search campaigns is illustrated in Fig.~\ref{fig:hist}, which indicates that the solutions below 700 MEUR are rare tail events with a proportion of approximately 0.40\% (84 out of 21{,}163 solutions of the merged campaign).

\section{Conclusions}
\label{sec:conclusions}

This study has shown that the total velocity increment of a multi-rendezvous sequence can be evaluated directly from an unordered target set, without determining either the visiting order or the rendezvous epochs. The resulting evaluator attains a mean relative error of about 5\% at less than one millisecond per sequence; it is therefore more accurate than the order-free analytical approximations of the same quantity and several orders of magnitude faster than optimizing each candidate sequence, and its error does not accumulate as targets are added. This combination of accuracy and speed is the main advantage of the model and the reason why it is effective inside a multi-spacecraft global search: the selection pressure acts on estimates that are close enough to the verified costs for superior target sets to survive, while the search can afford the large number of evaluations that exploring the solution space requires. Consistently, the best known solution of every fleet size considered here is obtained either directly or with a short refinement stage.

Two limitations qualify this conclusion. The model is accurate in the velocity-increment metric, whereas the mission cost amplifies the same relative error for sequences with a large velocity increment, so that the final ranking of the candidates still has to be established by precise optimization; and the model tends to underestimate sequences that are forced through a single high-cost leg under tight windows, which is exactly the situation in which an underestimated candidate is most likely to displace a better one during the search.

\section*{Acknowledgments}

This work was supported by the National Natural Science Foundation of China under Grant No.~12572407.

A generative artificial intelligence tool was used for language polishing, for translation, and as an aid in debugging the computer code. The author reviewed and edited the resulting content and takes full responsibility for the manuscript.

\bibliography{gtoc9_gnn_refs_clean}

\end{document}